\documentclass{emulateapj}

\shorttitle{Spectro-Perfectionism}
\shortauthors{Bolton \& Schlegel }
 
\begin{document}
 
\title{Spectro-Perfectionism: An Algorithmic Framework for Photon Noise-Limited
Extraction of Optical Fiber Spectroscopy}

\author{\mbox{Adam S. Bolton\altaffilmark{1,2}}}
\author{\mbox{David J. Schlegel}\altaffilmark{3}}

\altaffiltext{1}{Department of Physics and Astronomy, University of Utah,
115 South 1400 East, Salt Lake City, UT 84112, USA ({\tt bolton@physics.utah.edu})}
\altaffiltext{2}{Beatrice Watson Parrent Fellow,
Institute for Astronomy, University of Hawai`i,
2680 Woodlawn Dr., Honolulu, HI 96822, USA}
\altaffiltext{3}{Physics Division, Lawrence Berkeley National Laboratory,
Berkeley, CA 94720-8160, USA ({\tt djschlegel@lbl.gov})}

\begin{abstract}
We describe a new algorithm for the ``perfect'' extraction of
one-dimensional spectra from two-dimensional (2D)
digital images of optical fiber
spectrographs, based on accurate 2D forward modeling
of the raw pixel data.  The algorithm is correct for arbitrarily complicated
2D point-spread functions (PSFs), as compared to the traditional
optimal extraction algorithm, which is only correct for
a limited class of separable PSFs.  The algorithm results in
statistically independent extracted samples in the 1D
spectrum, and preserves the full native resolution of the
2D spectrograph without degradation.  Both
the statistical errors and the 1D resolution of the extracted
spectrum are accurately determined, allowing a
correct $\chi^2$ comparison of any model
spectrum with the data.
Using a model PSF similar to that
found in the red channel of the Sloan Digital Sky Survey
spectrograph, we compare the
performance of our algorithm to that of cross-section
based optimal extraction, and also demonstrate that our
method allows coaddition and foreground estimation
to be carried out as an integral part of the extraction step.
This work demonstrates the feasibility of current-
and next-generation multi-fiber spectrographs for
faint galaxy surveys even in the presence of strong
night-sky foregrounds.  We describe the handling of subtleties
arising from fiber-to-fiber crosstalk, discuss some of the
likely challenges in
deploying our method to the analysis of a full-scale survey,
and note that our algorithm could be
generalized into an optimal method for the rectification and
combination of astronomical imaging data.
\end{abstract}


\slugcomment{To be published in the PASP}

\maketitle

\section{Introduction}
Optical fibers offer a compelling advantage
for astronomical survey spectroscopy.  By allowing the light from targets
over a wide field of view on the sky to be rearranged into
a compact format and fed to any number of spectrographs in parallel,
they can provide a vast multiplex advantage over imaging
spectrographs.  For this reason, fiber technology has been adopted for a
number of major survey programs such as the Las Campanas
Redshift Survey (LCRS: \citealt{schech_lcrs}),
the Two-degree Field Survey (2dF: \citealt{col_2df}),
the Sloan Digital Sky Survey (SDSS: \citealt{york_sdss}), and
the recently initiated Baryon Oscillation
Spectroscopic Survey (BOSS: \citealt{sch_boss}) of the SDSS3 project.
The use of fiber spectrographs for faint-object spectroscopy has
however been restrained by concerns over throughput and systematic
limitations on the quality of subtraction of night-sky emission
foregrounds.  The spectroscopic extractions of the SDSS
multi-fiber instrument have established a high standard,
but significant systematic shortcomings remain.
Methods to characterize and partially remove
sky-subtraction residuals in SDSS fiber spectra can mitigate
the problem somewhat (e.g.\ \citealt{bol04, wild05}),
but do not substitute for a formally
correct sky-subtraction model.
The faintest spectroscopic galaxy surveys have
tended to make use of multi-slit imaging spectrographs
(e.g., \citealt{cow96, steidel03, dav03, gdds1, lef05}), along
with sky-subtraction techniques such as nod-and-shuffle \citep{cuil94, glaz01}
or B-spline-based modeling of the two-dimensional sky spectra in the
slits (e.g., \citealt{kel03}).
Nod-and-shuffle techniques in particular
are ill-suited to most fiber spectrographs, require at least
double the detector area, and furthermore reduce
the background-limited signal-to-noise by a factor of $1/\sqrt{2}$
due to their subtraction of data from data.

This paper outlines an algorithmic framework for the modeling
and extraction of optical and near-infrared astronomical fiber spectroscopy
to the limit of photon noise and native instrumental resolution.
By comparing our method with the standard techniques of optimal extraction
currently in wide use, we identify and resolve key systematic barriers
to ``perfect'' extraction.  As a result, multifiber spectroscopy emerges
as a clear and compelling technique for current and future generations of
faint-galaxy spectroscopic surveys, even well below the brightness of the
night sky at all wavelengths.  This algorithm will
deliver significant benefits to the reanalysis of the
original SDSS (hereafter SDSS1)
archive and to the ongoing analysis of the BOSS survey, both for
core redshift-survey goals and for projects that aim to select
rare emission-line objects from within the regions of the spectrum
dominated by OH line emission from the night sky
(e.g., \citealt{bol04, wil05, bol06, wil06, bol08}).

In considering this subject, we will make a clear
distinction between the problems of ``calibration'' and ``extraction'':
calibration is the description of the way in which any set of
astronomical and environmental stimuli translate into the
responses of the digital detectors (which we assume here to be pixellated
charge-coupled devices, or CCDs); extraction is the reconstruction of
particular stimuli from particular responses.
More strictly speaking, we view an ``extracted spectrum'' not
as a model for the flux of the observed source itself, but
rather as a properly calibrated one-dimensional
compression of the instrumental response to an observation
of that source.
When executed and reported correctly, an extracted spectrum permits
a statistically valid $\chi^2$ test of an input model spectrum
against the extracted pixels.
This paper specifically illustrates a method for carrying out
this sort of perfect extraction
assuming that perfect calibration is available.
We do not mean to trivialize the
problem of calibration, and
in our concluding remarks we will
discuss the relationship of our
extraction method to current and future spectroscopic calibration
regimes.

The paper is organized as follows.
Section~\ref{mod2d} frames the problem of extraction in terms of
image modeling, lays out the first part of our algorithm, and
compares its performance with that of standard extraction
techniques in terms of the quality of their respective
models to the raw 2D data.  Section~\ref{resol} addresses the
issue of resolution and covariance in the extracted spectra,
and presents the second part of our method, which
establishes optimal properties in both these regards.
Section~\ref{test4x3} presents a more realistic demonstration
of our algorithm on simulated data, illustrating several
additional strengths and subtleties of the method.
Finally, Section~\ref{dc} provides a discussion and
conclusions.

We will observe the following conventions in this paper.
Without loss of generality, we will assume that
spectroscopic traces run roughly parallel to CCD columns
(i.e., ``vertically''), with wavelength increasing with
row number and cross-sectional position increasing with
column number.  We will denote vectors in lowercase bold-face
type ($\mathbf{f}$) and matrices in uppercase bold-face type
($\mathbf{A}$).  We assume all errors have a Gaussian
distribution, and we assume no formal priors on fitted model
parameters.

\section{Extraction as image modeling}
\label{mod2d}

The current standard of quality for the extraction of optical fiber
spectroscopic data from digital images to 1D spectra is the
optimal technique described by \citet{hewett_85},
\citet{horne_86}, and \citet{rob86}, and subsequently
implemented in countless forms (e.g.,
\citealt{mar89, kin91, hall94, bac01, pisk02, cush04, ber05, zan05, dix07, bol07, cui08}).
This algorithm models the two-dimensional spectrum image
row by row with a minimum-$\chi^2$ scaling of the
cross-sectional profile.  The fitted amplitude is taken as
an optimal estimator of the 1D spectrum at the wavelength
corresponding to that row.
Optimal extraction as currently understood and practiced
has significant advantages, most notably in the increased
signal-to-noise ratio of the extracted spectrum in comparison
with the boxcar-aperture summing technique.  It also allows for model-based
flagging of pixels afflicted by cosmic-ray hits.  Optimal
extraction furthermore leads to statistically uncorrelated extracted
spectrum values, although this property is not preserved under
the rebinning that is required for the combination of multiple
exposures that are not precisely aligned at the raw-pixel level.

The most significant (and under-appreciated)
shortcoming of the optimal extraction method
is that it is only correct in the case where the
\textit{two-dimensional} image $I$ of a monochromatically illuminated
fiber (the point-spread function, or PSF)
is a separable function of column $x$
and row $y$ offset from the PSF centroid:
\begin{equation}
I (x, y) = I_x (x) I_y (y)
\end{equation}
If this condition does not hold---and it does not
hold for nearly all PSF models beyond a single Gaussian
ellipsoid whose principal axes are aligned with
the CCD rows and columns---then the cross-sectional
profile of the 2D spectrum is no longer fixed, but
rather depends upon the details of the input spectrum itself.
This is of limited concern if the spectrum is smoothly
varying with wavelength.
However if the spectrum involves many sharp features
such as the bright OH rotational emission lines in the
near infrared spectrum of the night sky,
then the output of the row-by-row optimal extraction algorithm
becomes ill-defined.  For example,
the cross section through the core of an emission line may have
a different shape than the cross section through the wing
of the line, yet the extraction will attempt to model them both
with a single average cross-sectional shape.  This will
degrade the resolution and signal-to-noise ratio of the extracted spectrum,
and may introduce a bias into the spectrum estimation.

The shortcomings of the row-wise optimal extraction method
can be overcome by modeling the 2D spectroscopic data image in a manner
that more accurately reflects the way in which an input spectrum
translates into photon counts in the CCD image.
For the case of a linear CCD detector (which we will assume), the system
calibration can be described as a matrix $\mathbf{A}$, whose elements
$A_{i \ell}$ describe the (noise-free) counts in CCD pixel $i$ for
a unit of monochromatic input at wavelength $\lambda_{\ell}$.  Note that
we are suppressing the natural two-dimensionality of the CCD by
allowing $i$ to index all column/row combinations.  We are also
suppressing any possible multi-fiber dependence; it is
straightforward to incorporate
this sort of dependence into the $\ell$-indexed dimension, and
we will in fact do so in \S\ref{test4x3} below.  The calibration matrix
$\mathbf{A}$ incorporates all the effects of wavelength calibration,
spectrum trace position, PSF shape and its dependence upon
position, flat-fielding effects, spectrophotometry, and
exposure time.  Note that
$\mathbf{A}$ will generally be a sparse matrix, since an input
at a given wavelength in a given fiber will only influence a relatively
small number of the full CCD pixels.
For the case of an input spectrum vector $\mathbf{f}$, the
observed CCD pixel count vector $\mathbf{p}$ will be given by
\begin{equation}
\label{f2p}
\mathbf{p = A f + n} ~,
\end{equation}
or in indexed notation
\begin{equation}
p_i = \left( \sum_{\ell} A_{i \ell} f_{\ell} \right) + n_i ~,
\end{equation}
where $\mathbf{n}$ is a pixel noise vector.
Extraction is then the reconstruction of $\mathbf{f}$
from $\mathbf{p}$ given a knowledge of $\mathbf{A}$.
Although extraction is non-trivial,
it is a linear problem.

Since the wavelength coordinate is continuous,
Equation~\ref{f2p} should most properly be written
in integral rather than matrix form.
However our interest is in \textit{solving} for $\mathbf{f}$:
inversion for an infinite number of differentially
spaced wavelength amplitudes from a finite number of pixels
and resolution elements is neither possible nor desirable.
We therefore restrict our attention to
a discrete set of wavelength sampling positions
$\left\{\lambda_{\ell}\right\}$.
The means of determining the most appropriate sampling
density for these positions will be addressed in more
detail further below; in general they may be either more closely or
more widely spaced than the pixel rows in the raw data.
However, if they are too closely spaced, then the solution
for $\mathbf{f}$ becomes degenerate.

\begin{figure*}[t]
\plottwo{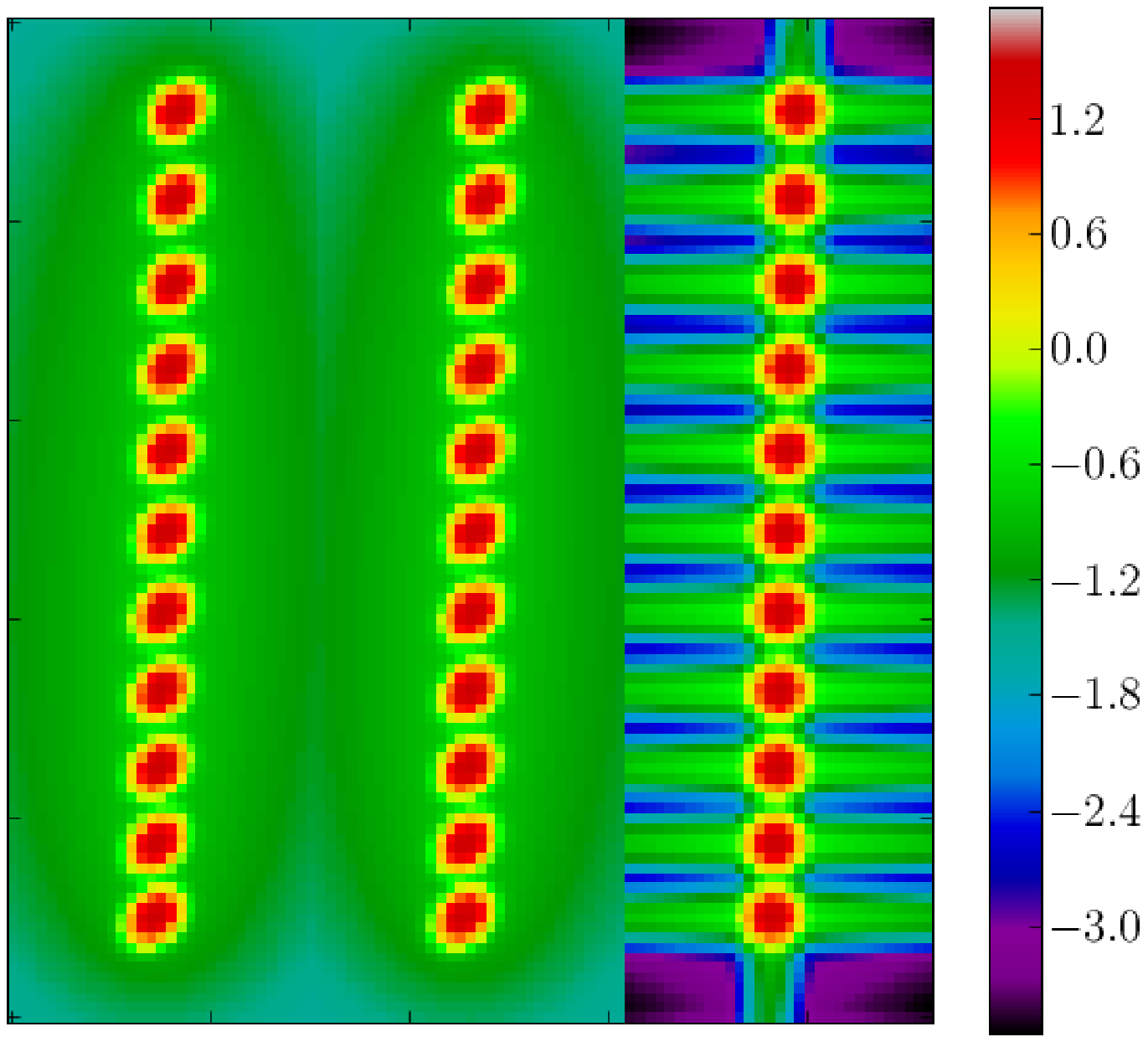}{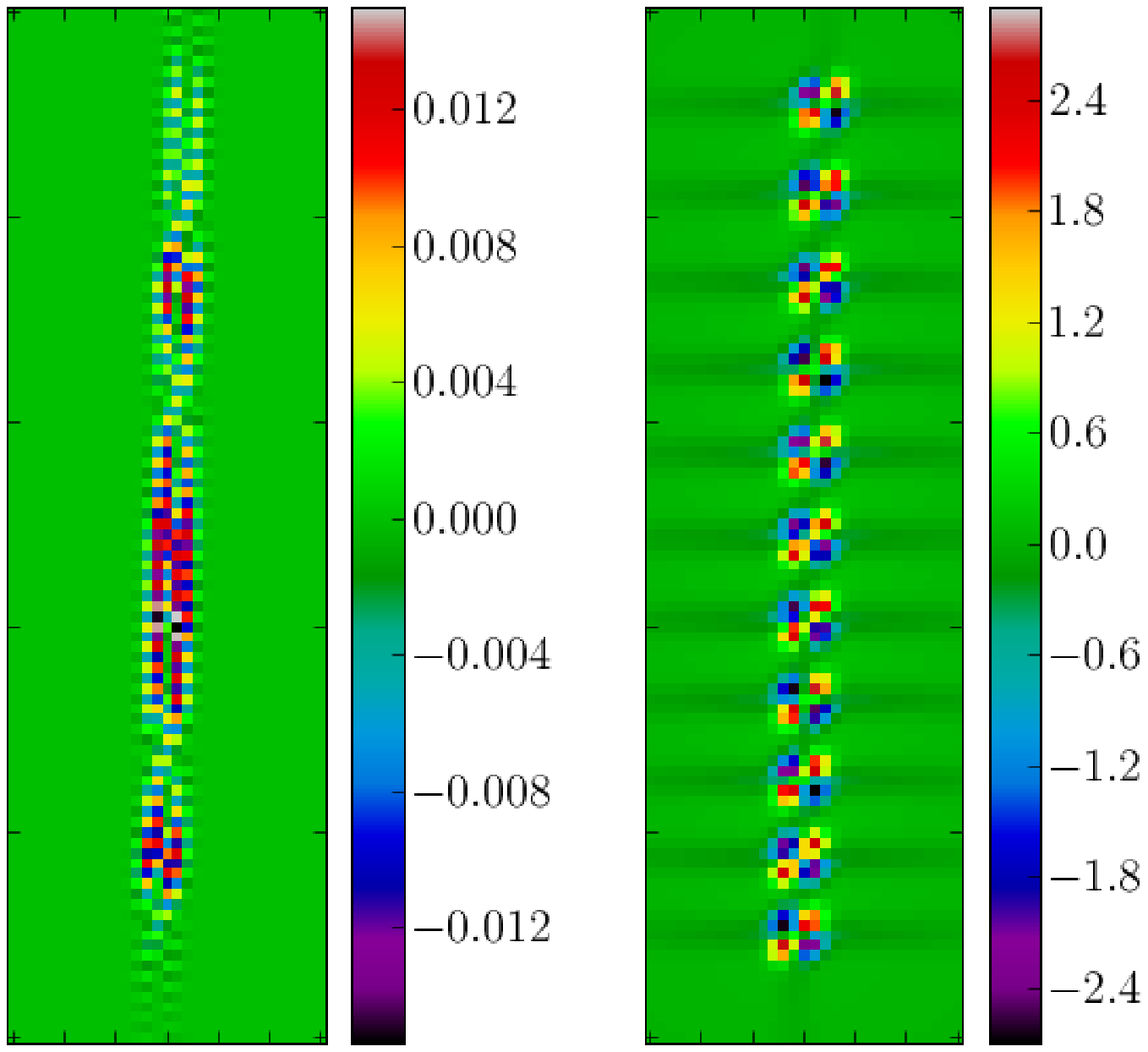}
\caption{\label{exmodel}
\textit{Left:} from left to right: simulated
noise-free emission-line image,
2D extraction model of simulated image, and
row-wise extraction model of simulated image.  Color
scaling is in units of the base-10 logarithm of the
pixel value, with the pixel values themselves scaled to
have an average value of unity across the entire image.
\textit{Right:} residuals resulting from the subtraction
of the 2d extraction model (left) and row-wise extraction model
(right) from the simulated emission-line image.
Color scale is in (non-logarithmic)
units of residual counts.  Image regions are 31$\times$101
pixels in size in all cases.}
\end{figure*}

Assuming Gaussian noise, the minimum-$\chi^2$ solution for
the input spectrum vector $\mathbf{f}$ from the data vector
$\mathbf{p}$ is
\begin{equation}
\label{solution1}
\mathbf{f} = \left(\mathbf{A^T} \mathbf{N}^{-1} \mathbf{A}\right)^{-1}
\mathbf{A^T} \mathbf{N}^{-1} \mathbf{p}~.
\end{equation}
Here, $\mathbf{N}$ is a pixel noise matrix,
with $N_{ij} = \left< n_i n_j \right>$.
We treat raw pixel errors as statistically independent, and thus the
noise matrix is diagonal.  Our model for the 2D raw pixel data is
simply given by $\mathbf{m} = \mathbf{A f}$.  Again, this requires
that the sampling points not be too closely spaced in wavelength,
so that $\mathbf{A^T} \mathbf{N}^{-1} \mathbf{A}$ does not
become non-invertible.

To demonstrate how this extraction method improves upon row-by-row
optimal extraction based on a fixed cross-sectional profile, we will
adopt the following illustrative fiber spectrograph PSF:
\begin{equation}
\nonumber
I(x,y) = {{(1-b)} \over {\sqrt{2 \pi}} \sigma} \exp \left[{{-r_{\mathrm{ell}}^2}
\over {(2 \sigma^2)}}\right] 
+ {{b \exp (-r / r_0)} \over {2 \pi r_0 r}}~.
\end{equation}
Here, $\sigma$ (in pixels) controls the size of a Gaussian core
to the profile (first term), and $r_0$ sets the characteristic size of the
profile wings (second term).  The parameter $b$ controls the fraction of the
total flux in the wing component.  The form of the wing component
is taken from a parameterization of the near-IR scattering seen
within the SDSS1 spectroscopic CCDs (J.E. Gunn, unpublished).
The coordinate $r = \sqrt{x^2 + y^2}$ is the radial offset in pixels on the CCD from
the center of the PSF spot.  The form $r_{\mathrm{ell}}$ in the
argument of the Gaussian term indicates that the core profile can
have an elliptical shape as determined by
\begin{equation}
r_{\mathrm{ell}} = \sqrt{q x^{\prime 2} + y^{\prime 2} / q}~,
\end{equation}
where $q$ is a minor-to-major axis ratio, and $(x^{\prime}, y^{\prime})$ are
related to the CCD column/row coordinates $(x,y)$ by a possible
rotation and translation.  For the wings, we adopt the values
$b = 0.1$ and $r_0 = 36$ pixels, roughly appropriate to the SDSS1
CCDs at a wavelength of 8500\,\AA\@.  For the core, we
take $\sigma = 1.1$ pixels, roughly the median characteristic
value for the SDSS1, and an ellipticity of $q=0.75$ with a major
axis position angle inclined at 45$^{\circ}$ relative to the
column/row axes.  This core ellipticity is somewhat more exaggerated
that what is seen in well-focused SDSS1 spectrograph exposures,
but will provide a good test of our method in the presence of
the astigmatic aberrations that can
arise in wide-field camera systems.

Generally speaking, the calibration matrix
$\mathbf{A}$ is determined by the combination of the fiber
PSF functional form, the dependence of its
parameters upon wavelength and position
(which can be derived from calibration frames illuminated
by gas discharge lamps
that have multiple discrete emission lines at known
wavelengths),
and the wavelength- and pixel-dependent throughput of the
system (which can be derived from calibration
frames uniformly illuminated by incandescent
lamps with approximately flat continuum spectra).
For simplicity, we will use CCD pixel rows as a proxy for wavelength,
neglect any variation of the PSF shape parameters with position,
and neglect any pixel-to-pixel sensitivity variations.
Assuming that all these effects can be calibrated successfully,
they can be incorporated into the approach that
we describe.  We also neglect the problem of cosmic rays here,
although a model-based extraction such as ours is ideally suited to the
iterative masking of cosmic-ray pixels in the raw data
based upon their statistical disagreement with the best extraction
model.  Finally, we ignore the possibility of a large-scale
scattered-light component in the image, but the incorporation of
this phenomenon (as well as those of imperfectly subtracted
instrumental bias and dark-current levels)
into our 2D modeling would be straightforward.

The most stringent test of any extraction is provided by
unresolved emission lines, which show the sharpest
real features possible in the data.  To test our method
alongside the standard row-wise extraction,
we therefore create a noise-free spectrograph-like image consisting of
11 emission lines of equal flux spaced vertically over
80 pixels within a 31$\times$101 pixel image.  We include
a slight tilt of the spectral trace relative to the vertical,
and incorporate a mild rate of change in the spacing
of the lines so as to ensure a range of sub-pixel positions
for the line centers.  Each pixel value is computed by integrating
the profile with 4$\times$4 sub-sampling.
We incorporate a small core radius in the denominator of
the scattering wing term to avoid numerical difficulties with the
(integrable) singularity.  This simulated image is shown in
the leftmost panel of Figure~\ref{exmodel}.

Next we perform an extraction of this model image using both
the standard row-wise, cross-section based optimal extraction
and our 2D modeling method.  For the 2D extraction, we
compute a basis set of 101 PSFs centered on the trace position
in each of the 101 image rows.  Note that the ``emission lines''
of the simulated image will in general not be coincident with
these basis functions, but will instead be offset from the
nearest basis PSF by some sub-pixel amount in y.  For the row-wise
extraction, we sum all 101 basis functions with equal weight to
generate the image of a flat continuum spectrum, and take the normalized
cross-sectional shape of this spectrum in each row as the
basis profile for the extraction of that row.  Thus both the
2D and row-wise extractions will have an equal number of
free extraction parameters, with the same sampling interval
between them.  Giving equal weight to all pixels, we compute
the linear least-squares set of amplitudes that multiply the
basis functions of each extraction method to best
reproduce the simulated image.  The results are shown in
Figure~\ref{exmodel}.  The difference between the quality
of the two models can be seen immediately, with the row-wise
extraction model unable to capture the variation of the
cross sectional shape with spectrum that comes from the ellipticity
of the PSF core and the presence of the scattering wings
acting in concert with the sharp spectral features.
Figure~\ref{exmodel} also shows the residuals of subtraction
of the two extraction models from the simulated image.
The residuals of the 2D method have a scale roughly 200$\times$
smaller than the residuals of the row-wise method.

We can make a quantitative comparison of the relative quality of the
two extraction models to the simulated image in terms
of the following two figures of merit:
\begin{eqnarray}
D_{\mathrm{sq}} &=& \left( \sum_i (p_i - m_i)^2 \left/
\sum_i p_i^2 \right. \right)^{1/2} ~, \\
D_{\mathrm{ab}} &=& \sum_i | p_i - m_i | \left/ \sum_i | p_i | \right. ~.
\end{eqnarray}
Here, the $p_i$ are the raw pixel data values, and the $m_i$
are the 2D extraction model values in those pixels.
Both of these figures will be equal to zero in the case of a
perfect model, and equal to 1 for a model that is identically zero
(i.e., no model at all).  For the case described above, we find
$D_{\mathrm{sq}} = 0.0006$ and $D_{\mathrm{ab}} = 0.0008$ for the
2D model, and $D_{\mathrm{sq}} = 0.12$ and $D_{\mathrm{ab}} = 0.20$
for the row-wise model, again confirming a factor of $\approx$200
improvement with the 2D model.

Before proceeding further, we address the question of appropriate
sampling.  While the sampling of the row-wise extraction is naturally
limited to one sampling point per row, the sampling of the 2D
extraction method can easily be adjusted to be either finer
or more coarse than this.  The residuals seen above after subtracting
the 2D model
to the simulated emission-line image, although very small compared
to those of the row-wise model, nevertheless exceed the numerical
precision of the calculation, and are due to the misalignment between the
extracting basis and the input emission lines in the
``wavelength'' direction.  We might reasonably improve upon the extraction
model by adopting a more finely-spaced basis set.
If the typical fiber spectrograph PSF were purely band-limited,
then the appropriate sampling could be chosen via the Nyquist
criterion.  However, fiber spectrograph PSFs are primarily determined
not by a diffraction limit, but rather by the convolution
of a sharp ``top-hat'' fiber image with a set of optical aberrations,
and thus the determination of an appropriately critical
extraction sampling must be made according to more empirical factors.
For the particular case under consideration here, we have tested the quality
of the 2D model to the simulated image for sampling densities ranging from
0.5 to 2 basis functions per row in the 2D image.
Both $D_{\mathrm{sq}}$ and  $D_{\mathrm{ab}}$ improve with increasingly
fine sampling until about 1.3 basis functions per row, where they
attain a value of a few $\times 10^{-5}$.  With finer sampling beyond
this point, the model becomes worse as the matrix
to be inverted approaches
singularity.  In practice, the appropriate sampling density will
be determined by the details of the spectrograph PSF and dispersion,
and consequently by the spectral resolution as defined
in the following section.

\section{Resolving the resolution}
\label{resol}

We have demonstrated the fidelity of our 2D extraction model to the
two-dimensional pixel data, but what of our extracted spectrum?
Naively, $\mathbf{f}$ from Equation~\ref{solution1}
is our 1D spectrum estimator, and the matrix
$\mathbf{C} \equiv (\mathbf{A^T} \mathbf{N}^{-1} \mathbf{A})^{-1}$
is the covariance matrix of its individual values.
But all is not well: Equation~\ref{solution1} not only extracts
the spectrum, it deconvolves the spectral resolution.
The instability of the deconvolution manifests itself
in a significant ``ringing'' in the extracted spectrum,
along with large correlations and anti-correlations between
extracted pixels.  Both of these are decidedly undesirable
features in a spectrum.
The way to remedy this situation is the second key element of
our extraction method: to
\textit{re-convolve} our de-convolved spectrum back to
the same resolution as the raw data.

Consider the inverse covariance matrix $\mathbf{C}^{-1}$ of
the deconvolved spectrum basis:
\begin{equation}
\label{covar}
\mathbf{C}^{-1} = \mathbf{A^T} \mathbf{N}^{-1} \mathbf{A} ~.
\end{equation}
This matrix is symmetric and band-diagonal, and all of its entries are
non-negative (assuming that there is no way for any input spectrum
to \textit{subtract} counts from the CCD)\@.  Now take the 
unique non-negative matrix square root of $\mathbf{C}^{-1}$
to find a symmetric matrix $\mathbf{Q}$ such that
\begin{equation}
\label{msqrt}
\mathbf{C}^{-1} = \mathbf{Q Q}~.
\end{equation}
This may be done by determining the eigenbasis of $\mathbf{C}^{-1}$,
taking the element-wise square root of the diagonal
matrix of its eigenvalues (which will all be positive
since $\mathbf{C}^{-1}$ is positive definite),
and transforming this new diagonal matrix back using the unitary matrix
that relates the eigenbasis to the original basis.
Equation~\ref{msqrt} would still hold if we applied any
arbitrary set of sign-flips to the square-root-eigenvalue
matrix (leading to other matrix square roots besides the
unique non-negative one), but these would introduce undesired negativity
into $\mathbf{Q}$.

Next, define a normalization
vector $\mathbf{s}$ through
\begin{equation}
\label{sdef}
s_{\tilde{\ell}} = \sum_{\ell} Q_{\tilde{\ell} \ell}~,
\end{equation}
a matrix $\mathbf{R}$ through
\begin{equation}
\label{rnorm}
R_{\tilde{\ell} \ell} = s^{-1}_{\tilde{\ell}} Q_{\tilde{\ell} \ell}
~~\mbox{(no sum)},
\end{equation}
and a diagonal matrix $\mathbf{\widetilde{C}}^{-1}$ with entries
given by
\begin{equation}
\widetilde{C}_{\tilde{\ell} \tilde{\ell}}^{-1} = s_{\tilde{\ell}}^2~.
\end{equation}
By construction, we now have
\begin{equation}
\label{newcovar}
\mathbf{C}^{-1} = \mathbf{R^T} \mathbf{\widetilde{C}}^{-1} \mathbf{R}~,
\end{equation}
and consequently
\begin{equation}
\mathbf{\widetilde{C}} = \mathbf{R} \mathbf{C} \mathbf{R^T}~.
\end{equation}

Note the complete analogy between Equations \ref{covar} and~\ref{newcovar}.
The matrix $\mathbf{R}$ defines a transformation from the
deconvolved spectrum to a ``reconvolved'' spectrum for which
(1) the extracted 1D pixels are statistically independent of
one another (since the matrices $\mathbf{\widetilde{C}}^{-1}$
and $\mathbf{\widetilde{C}}$ are
diagonal), and (2) the blurring of input spectra in wavelength
from the deconvolved basis
is statistically equivalent to the blurring imposed by the
actual 2D spectrograph matrix $\mathbf{A}$ on the physical
input spectrum.  It is in this sense
that we can re-convolve with the ``same'' resolution as
was inherent in the 2D data.  Our extracted
1D spectrum is then
\begin{equation}
\mathbf{\widetilde{f} = R f}~,
\end{equation}
with uncorrelated errors described by the diagonal covariance
matrix $\mathbf{\widetilde{C}}$ and undegraded resolution
described by the matrix $\mathbf{R}$.
We may think of this reconvolved spectrum as a
model for what \textit{would have been observed} by a truly
one-dimensional spectrograph with the same resolution as
our two-dimensional CCD spectrograph.
Since the resolution of the extracted spectrum is
precisely characterized by the matrix $\mathbf{R}$, we can
convolve any theoretical model for the input 1D
spectrum with this matrix in order to compare to the data
and compute a statistically correct $\chi^2$ value.
The sense of the normalization
imposed by Equation~\ref{rnorm} is to describe the elements
of $\mathbf{\widetilde{f}}$ as a weighted sum over the
elements of the theoretical input spectrum, with the sets
of weights individually summing to 1.

\begin{figure}[b]
\plotone{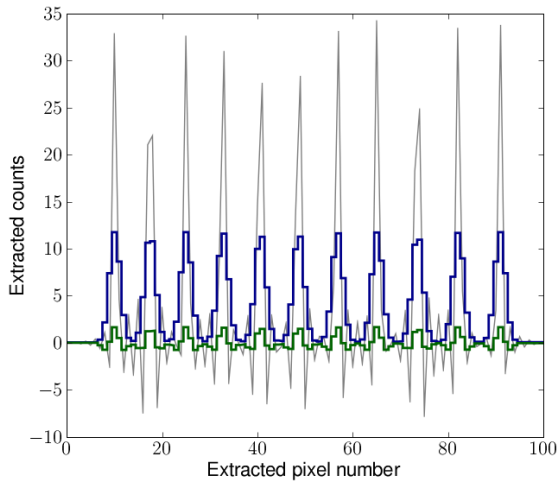}
\caption{\label{spec1}
Extracted spectra of simulated noise-free emission-line image.
\textit{Thin black line:} deconvolved spectrum
from the 2D modeling extraction method.  Note significant ringing.
\textit{Thick blue line (upper of the two thick lines,
rendered with steps):}
deconvolved spectrum from 2D modeling extraction
reconvolved to the native spectrograph resolution using the
resolution matrix $\mathbf{R}$ defined in the text.
\textit{Thick green line (lower of the two thick lines,
rendered with steps):}
10$\times$ the difference between the
reconvolved 2D-model extracted spectrum and the row-model extracted
spectrum. ``Upward-peakiness'' at the position of the emission lines
indicates that the 2D-extracted spectrum has higher resolution
than the row-extracted spectrum.}
\end{figure}

Figure~\ref{spec1} illustrates the deconvolved and reconvolved
spectra that result from the extraction of the simulated
emission-line spectrum of \S\ref{mod2d}.  Note that the
ringing in the deconvolved spectrum is completely absent
from the reconvolved spectrum.
The peaks in the reconvolved
spectrum represent the true undegraded resolution of the
2D spectrographic data.
The values in the deconvolved
spectrum are highly correlated with one another, whereas
the values in the reconvolved spectrum are completely uncorrelated.
Also shown is a representation
of the difference between the extracted spectra using
the 2D and row-wise methods.  Although the two spectra are
sufficiently similar that their differences cannot be seen in
a direct plot, they differ at the few percent level,
which is significant for application to the extraction of faint galaxy
spectra in the presence of strong night-sky line emission.  Also note that
the resolution of the 2D-extracted spectrum is higher than the resolution
of the row-extracted spectrum.

In defining the resolution with which to reconvolve our
deconvolved extracted spectrum, we have explicitly chosen a form that
results in an exactly diagonal sample covariance matrix in
the reconvolved basis.  Additionally, as a consequence of the
band-diagonal nature of $\mathbf{C}^{-1}$, our $\mathbf{R}$ matrix is
also essentially band-diagonal, corresponding to a localized
``line-spread function'' in the extracted 1D spectrum
as seen in Figure~\ref{spec1}.
It is nevertheless worth remembering that other
choices for $\mathbf{R}$ are available.  For instance, there may be applications
for which it is desirable to have a simpler parameterized form
for the resolution than will in general result from
the procedure outlined above.  The parameters of this
form could be optimized so as to make the off-diagonal correlations
in the reconvolved basis sufficiently small while still non-zero.

\section{A more interesting test}
\label{test4x3}

\begin{figure*}[t]
\plotone{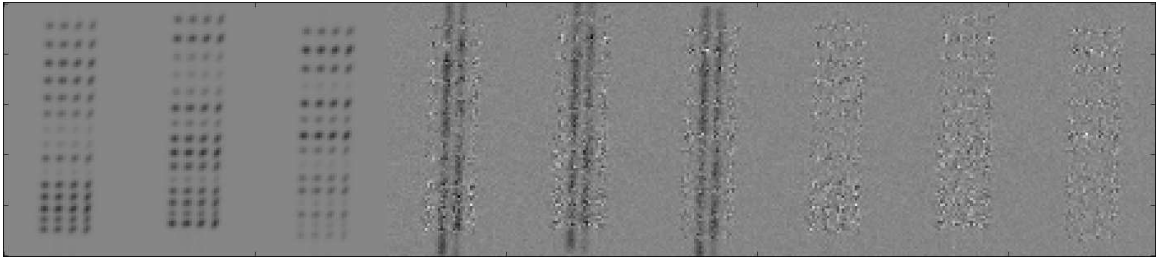}
\caption{\label{modelframe}
Simulated multifiber, multi-exposure spectroscopic data,
including noise, flexure, a non-uniform PSF, and ``sky''.
\textit{Left:} Three ``exposures'' of four
fiber spectra, including simulated flexure and sky-spectrum
variability.  Object spectra are included in central two fibers
in each set, but are too faint to see directly.  \textit{Center:}
same as left, after subtraction of extraction model for the sky
component in each exposure, with gray-scale stretched by a factor
of 40 to reveal the traces of the object spectra.  \textit{Right:}
As in center, but after subtraction of extraction model for the
object spectra as well.  When scaled by the pixel errors, these residuals
are consistent with a reduced $\chi^2$ of unity.  Each ``exposure''
is 51$\times$101 pixels in size.}
\end{figure*}

Before taking up the longer-term challenge of implementing
our algorithm on actual survey spectroscopy,
we illustrate its power with a more realistic simulation
than the simple case above.  In particular,
we now include the following effects, all of which will
be found in real survey data:
\begin{enumerate}
\item Noise,
\item A varying 2D fiber PSF shape,
\item Overlap (``cross-talk'') between the 2D images of
neighboring spectra on the CCD detector of a multifiber spectrograph,
\item Multiple exposures with relative movement (``flexure'')
of the fiber images on the CCDs between them, and
\item A night-sky foreground that may vary between exposures
and must be modeled and subtracted to reveal a much fainter object
spectrum of interest.
\end{enumerate}
We use nearly the same PSF model as above, but now generate four ``spectra''
on the same image, separated by 5.7 pixels from one another in
the horizontal direction.  This separation is somewhat smaller
than the $\approx$6.2 pixel separation between neighboring
spectra in the SDSS1, and therefore leads to greater
cross-talk.  The Gaussian core of each fiber is given
a different ellipticity (ranging from 0.9 to 0.6)
and position angle (ranging between 15$^{\circ}$ and 60$^{\circ}$ from
the $x$ pixel direction), to mimic optical variations and
fiber heterogeneities.
The central two spectra are taken as ``object''
spectra, and the outside two are ``sky'' spectra.
We also generate three ``exposures'' of these four spectra,
shifting the fiber images relative to the CCD grid
in each case to simulate spectrograph flexure.
We generate a ``sky'' spectrum of 15 emission lines
at increasing separation over the range of the spectra.
In each of the three exposures, the amplitudes of these individual
sky lines are randomized, and the sky realization for each
exposure is projected through all four ``fibers'' in that
image.  The positions of the sky lines in wavelength are fixed
across all exposures.  Next, two ``object'' spectra are
projected through the central two fibers of all three exposures
and added to the image of the sky spectra.  The relative fluxes
are taken so that, in the fibers containing both object and sky,
the total sky counts exceed the total
object counts by a factor of 20.  We use an extraction sampling
density of 1.2 basis functions per CCD row.  The noise level
is set by assigning $10^4$ statistical sky counts (i.e., photons)
per spectrum per sampling point in total across all three exposures.
This implies 500 total object counts per sampling point, and
an approximate background-limited signal-to-noise ratio of
5 per extraction pixel for the objects.  We also include a ``read-noise''
term of 5 counts per pixel RMS\@.  We use the statistical noise level
to generate a noise-image realization that we add to the sky
and object data.  The resulting three sets of four spectra are seen
in the leftmost third of Figure~\ref{modelframe}.

The full power of the 2D modeling extraction now
becomes apparent.  We construct a generalized calibration
matrix, involving input spectra of both the night sky (one
for each exposure) and the science
target objects (one for each object).
In each exposure, the sky basis projects into the images of all fibers---sky
and object alike---while the object bases project
only into the images of the object fibers.
The multiple exposures of each object spectrum are
extracted in a single step: there is no need for
the separate steps of registration, extraction, and coaddition
of individual frames.
Similarly with the sky spectra, we extract a single
sky spectrum from all fibers of a given exposure.  Within
the object fibers, we extract the sky and object spectra in sum
together.  There is no explicit sky-subtraction step, but rather
a sky--object decomposition that is an integral part of the extraction
into individual 1D spectrum components.  All at once, we model and
extract three skies---one for each exposure---and two object spectra.
In summary: extraction, coaddition, and sky subtraction are all subsumed into
a single image-modeling operation.
The results of this modeling can be seen in the
second and third parts of Figure~\ref{modelframe}, illustrating
the sky--object decomposition and the ``photon-noise'' limited
quality of the sky$+$object model to the three 2D exposures.

We note that the all-in-one sky modeling and extraction is
necessary in order to model and decompose the sky spectrum from
the object spectra \textit{in the deconvolved frame}, upstream from the
fiber-to-fiber PSF variations that would lead to systematic
errors in a traditional modeling and subtraction
of the sky from extracted and resolution-convolved spectra.
If necessary, an accounting for any spatial variations of the
sky brightness over the telescope focal plane could be
built in by way of additional linear sky components.
The combination of multiple object spectra at the time
of extraction, on the other hand, is
not strictly necessary.  Sky-subtracted object spectra from
the individual exposures could be combined together in a subsequent
step, to allow for the non-linear determination of spectrophotometric
variations between the frames.  However, once these variations
are determined, the proper combination of these multiple frames would
resemble a second extraction, with the concatenation of their
individual $\mathbf{R}$ matrices forming the new $\mathbf{A}$
matrix (in the notation of \S\ref{mod2d}).

The presence of crosstalk between neighboring spectra, and
between the object spectra and the skies, makes the
determination of the 1D resolution matrix somewhat more challenging.
Consider the full covariance matrix $\mathbf{C}$ and
inverse covariance matrix $\mathbf{C}^{-1}$ in the
deconvolved extracting basis:
they consists of multiple blocks, with each block describing
the statistical coupling between different extracted spectra.
The matrix $\mathbf{C}^{-1}$ is band diagonal in all blocks, but
it has non-zero elements in the off-diagonal blocks coupling
different extracted spectra to one another.
Assuming the spectral cross-talk on the CCD is from spectra
that are otherwise unassociated with one another,
we do not simply want to take the matrix square root of $\mathbf{C}^{-1}$
to define our resolution, since the resolution defined in
this way would mix extracted spectra with one another.
A possible solution which we adopt here is to invert $\mathbf{C}^{-1}$
to obtain $\mathbf{C}$, zero out all the entries in the
off-diagonal blocks of $\mathbf{C}$, re-invert,
and define the resolution in terms of the square
root of the resulting matrix.
To make this explicit with block matrix
notation for a case of two extracted spectra:
\begin{eqnarray}
\mathbf{C} &=& \left(\mathbf{C}^{-1} \right)^{-1}
= \left[ \begin{array}{cc}
\mathbf{C}_{11} & \mathbf{C}_{12} \\
\mathbf{C}_{21} & \mathbf{C}_{22}
\end{array} \right] ~,  \\
\not \!\! \mathbf{C} &\equiv&
\left[ \begin{array}{cc}
\mathbf{C}_{11} & \mathbf{0} \\
\mathbf{0} & \mathbf{C}_{22}
\end{array} \right]~, \\
\not \!\! \mathbf{C}^{-1} &=&
\left[ \begin{array}{cc}
[\mathbf{C}_{11}]^{-1} & \mathbf{0} \\
\mathbf{0} & [\mathbf{C}_{22}]^{-1}
\end{array} \right] = \mathbf{QQ}~,
\end{eqnarray}
and proceed to define the resolution matrix $\mathbf{R}$ in
terms of this $\mathbf{Q}$ as indicated by Equations
\ref{sdef} and \ref{rnorm}.  Note that
$[\mathbf{C}_{11}]^{-1} \neq \mathbf{C}_{11}^{-1}$
and $[\mathbf{C}_{22}]^{-1} \neq \mathbf{C}_{22}^{-1}$
due to the non-zero off-diagonal blocks in the
constructed inverse covariance matrix $\mathbf{C}^{-1}$.

It can be shown that
by this definition, the resolution matrix will diagonalize
$\mathbf{C}$ \textit{within its diagonal sub-blocks}
so that extracted samples will be statistically independent
of one another within each spectrum.
There will in general be non-zero correlations
between the extracted
samples of different spectra with one another,
but this is unlikely to be important for spectra that are otherwise
unrelated.  Note that if the problem were rather to extract
the spectra of an integral-field spectrograph, with
cross-talk between fibers that were also adjacent to one another
on the sky, then it would perhaps be desirable instead to have a resolution
matrix that mixes spectra with one another but provides
full statistical independence of samples both within a single
spectrum and among neighboring spectra.

\begin{figure}[b]
\plotone{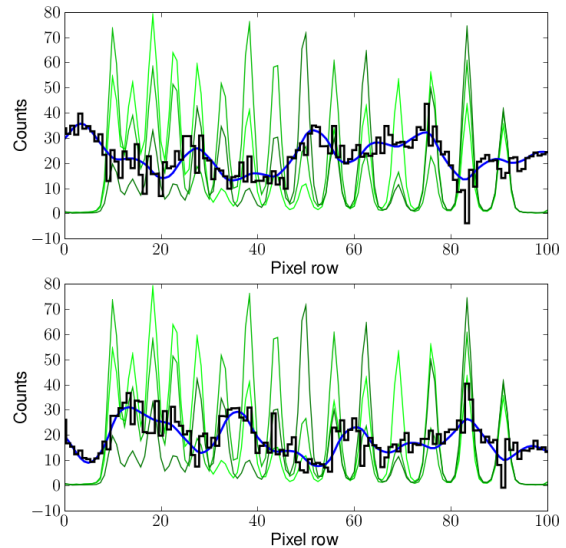}
\caption{\label{multi4x3}
Extracted spectra of two simulated objects from
multiple exposures as described in \S\ref{test4x3}
and depicted in Figure~\ref{modelframe},
together with extracted ``sky'' spectra.  Black lines (rendered
with steps) show the extracted spectra,
while blue lines (rendered smoothly and tracing the black lines)
show the input spectra convolved with the 1D resolution.
Green lines of varying shades (thinner, with higher peak
values, and identical in both plots) indicate the extractions of
the three different realizations
of the ``sky'' spectrum in the three individual exposures,
divided by a factor of 20 for display purposes.}
\end{figure}

Computing the resolution of our simulated multi-fiber, multi-exposure
set in the manner outlined above, and re-convolving the extracted object and sky spectra,
we obtain the results shown in Figure~\ref{multi4x3}.  The sky spectra
are scaled down by a factor of 20 for display purposes.  When weighted
by the error estimates, the extracted spectra have a reduced $\chi^2$ of
approximately unity relative to the resolution-convolved input spectra.
This demonstrates the suitability of our approach for the
extraction of faint galaxy spectra in the presence of high and
sharply wavelength-dependent night-sky foregrounds.

\section{Discussion and conclusions}
\label{dc}

The extraction algorithm that we have described does not
come without a price.  In the presence of
fiber-to-fiber cross-talk, the standard row-wise
optimal extraction will couple extracted amplitudes between
fibers in a single row, leading to a banded matrix of
dimension equal to the number of fibers that must be inverted;
this process must then be repeated for each row in each exposure.
Our 2D modeling extraction, on the other hand, couples
extracted amplitudes between fibers, wavelengths, and exposures.
Thus the matrix to invert for the solution set of spectra
has sides of dimension equal to
\begin{equation}
N_{\mbox{spectra}} \times N_{\mbox{samples per spectrum}} \times N_{\mbox{exposures}}~.
\end{equation}
For one SDSS1 pointing, this would correspond to
320 fibers (one of the two spectrographs), approximately 4000 sampling points,
and three exposures: i.e., a square matrix nearly 4 million
to a side.  With a brute-force approach this matrix
could never be stored, let alone inverted,
with any hardware of the present or foreseeable future.  The way
forward to determining the extracted spectra will no doubt lie in
exploiting the sparseness of the inverse covariance matrix
to reduce storage and computation, and to apply an iterative
method such as the conjugate gradient to solve for the extracted
spectra.  To determine the resolution with which
to reconvolve the extracted spectrum, which formally
requires the inversion of the full inverse covariance matrix $\mathbf{C}^{-1}$ ,
the practical solution will be to invert a sufficient subspace
of $\mathbf{C}^{-1}$ surrounding each spectrum (or subsegment
of a spectrum) to define
an acceptably accurate approximation to the desired resolution matrix.
The exact requirements will depend upon the degree of
cross-talk between neighboring fibers in the spectrograph under consideration.
Even with these strategies, a usable implementation
of our algorithm for real multifiber survey data may
require high-performance parallel computing, depending on the computational
expense of the necessary matrix-element calculations.

Most immediately, we plan to implement the strategy outlined in
this paper to the extraction of spectra from the BOSS Survey.
We also plan to conduct a reanalysis of the SDSS1 archive,
to provide the definitive version of this important spectral database
with the best possible extracted resolution, signal-to-noise, and
foreground subtraction.  These techniques also offer promise for
the upcoming Apache Point Observatory Galaxy Evolution Experiment
(APOGEE: \citealt{ap_apogee})
of the SDSS3, a high-resolution, near-infrared multifiber survey of
red giant stars in our Galaxy that aims to constrain their
evolutionary history as traced by multiple chemical abundances.
This extraction strategy will also form a key part of the
technical feasibility of the BigBOSS survey \citep{sch_bigboss},
which proposes to use a 5000-fiber spectrograph fed by a $3^{\circ}$
field-of-view focal plane positioner system on a 4m-class telescope
to measure the baryon acoustic scale and redshift-space distortions
over the redshift range $0.2 < z < 3.5$.  Although the
implementations for these different surveys will differ in
detail, we believe that the software engine for the core extraction
computations can be written in a general-purpose form.
The application of this technique to slit spectroscopy may be
possible as well, although the preservation of object spatial
information by the slit makes the problem substantially more
complicated.

In all cases, the full power of this extraction can only be realized
with sufficiently accurate calibration.  The current standard
calibrations for fiber spectroscopy are exposures of uniform
spatial illumination by
flat-spectrum incandescent lamps (``flats'') and
multi-emission line
gas-discharge lamps (``arcs'').  Within the framework of
our extraction algorithm, the former will determine the relative
sensitivity of the individual fibers and pixels, and the latter
will determine the spectrograph PSF shape and position as a function
of illuminating wavelength in each fiber.  Assuming there are no
systematic offsets between the calibrations and the science exposures,
and assuming that the variation of the spectrograph PSF is smooth
enough with wavelength to be well-sampled by the arc frames
(which are sparsely distributed in wavelength),
these calibrations should contain sufficient information to
determine the calibration matrix $\mathbf{A}$.
We may proceed by extracting the arcs and flats together,
with each one described by a single
spectrum projected through an initial guess for $\mathbf{A}$,
and then optimize the parameters of $\mathbf{A}$ by
non-linear iteration so as to improve the quality of the 2D
extraction models to convergence.
A more direct and ambitious approach to the determination
of the calibration matrix would be to illuminate the facility
calibration screen with either a high-wattage monochrometer
or a tunable laser (c.f.\ \citealt{st06, cra09}), and to step the illumination source
through wavelength in subsequent
exposures so as to map out $\mathbf{A}$ explicitly.
In practice, with the exception of spectrograph systems that
are very stable thermally and mechanically, there is likely
to be some shifting of the fiber positions and focus on the
CCDs between the calibration and science frames.
In this case, ``tweaks'' to the parameters of
$\mathbf{A}$ will be derived from the shapes and positions of the
fiber traces and night-sky emission lines in the science frames.
Finally, we note that the calibration may be significantly
improved by incorporating all available knowledge about the
optical design of the telescope and instrument, rather than
treating the system as a black box to be specified entirely
by empirical calibration data.

Putting aside the computational challenges that arise
from the consideration of continuously two-dimensional input data,
the method we have described can also be generalized into an
optimal recipe for the rectification and combination of multiple CCD imaging
exposures: i.e., taking the $\mathbf{f}$ of \S\ref{mod2d}
to be a \textit{two-dimensional} image model to
be extracted from the data.
As with the spectroscopic application, the resulting extractions would
have optimal resolution, statistically independent extracted image samples,
and a definition in terms of the optimization of a well-motivated
scalar objective function describing the quality of a model fit
to all of the individual exposures.  The implementation would be
non-trivial, but the benefits could be great.  A significant
challenge on the calibration side
is that the details of the imaging PSF, which must be known,
will depend upon the spectral energy distributions of the
imaged objects, which will in general be unknown.

In summary, we have demonstrated a new spectrum extraction algorithm
for optical and near-infrared astronomical fiber spectroscopy.
Given sufficiently accurate calibration, this method can extract
spectra to the statistical noise limit in the presence of arbitrarily
complicated point-spread functions and arbitrarily high and wavelength-varying
foregrounds.  The extracted spectra have optimal and precisely
quantified resolution and signal-to-noise, along with
statistically uncorrelated pixels,
for any number of sub-exposures in combination together.
As such, statistically accurate $\chi^2$ tests may be made
between the extracted data and theoretical models
of the input object spectrum.
This algorithm represents a fundamental
improvement upon the current state-of-the-art methods in use for
the extraction of fiber spectroscopy, and thus motivates a
serious and positive reevaluation of the promise of fiber-fed spectrographs for
next-generation ground-based faint-object surveys.

\acknowledgments
The authors wish to thank Scott Burles, Julian \mbox{Borrill},
Robert Lupton, David Hogg, Sam Roweis, and Michael Blanton for valuable
comments and discussions of this subject.  DJS acknowledges the support of
the Director, Office of Science, of the U.S. Department of Energy
under Contract No.\ DE-AC02-05CH11231.

\newpage


\begin{thebibliography}{36}
\expandafter\ifx\csname natexlab\endcsname\relax\def\natexlab#1{#1}\fi

\bibitem[{{Abraham} {et~al.}(2004)}]{gdds1}
{Abraham}, R.~G., {et~al.} 2004, \aj, 127, 2455

\bibitem[{{Allende Prieto} {et~al.}(2008)}]{ap_apogee}
{Allende Prieto}, C., {et~al.} 2008, Astronomische Nachrichten, 329, 1018

\bibitem[{{Bacon} {et~al.}(2001){Bacon}, {Copin}, {Monnet}, {Miller},
  {Allington-Smith}, {Bureau}, {Carollo}, {Davies}, {Emsellem}, {Kuntschner},
  {Peletier}, {Verolme}, \& {de Zeeuw}}]{bac01}
{Bacon}, R., {et~al.} 2001, \mnras, 326, 23

\bibitem[{{Bershady} {et~al.}(2005){Bershady}, {Andersen}, {Verheijen},
  {Westfall}, {Crawford}, \& {Swaters}}]{ber05}
{Bershady}, M.~A., {Andersen}, D.~R., {Verheijen}, M.~A.~W., {Westfall}, K.~B.,
  {Crawford}, S.~M., \& {Swaters}, R.~A. 2005, \apjs, 156, 311

\bibitem[{{Bolton} \& {Burles}(2007)}]{bol07}
{Bolton}, A.~S., \& {Burles}, S. 2007, New Journal of Physics, 9, 443

\bibitem[{{Bolton} {et~al.}(2006){Bolton}, {Burles}, {Koopmans}, {Treu}, \&
  {Moustakas}}]{bol06}
{Bolton}, A.~S., {Burles}, S., {Koopmans}, L.~V.~E., {Treu}, T., \&
  {Moustakas}, L.~A. 2006, \apj, 638, 703

\bibitem[{{Bolton} {et~al.}(2004){Bolton}, {Burles}, {Schlegel}, {Eisenstein},
  \& {Brinkmann}}]{bol04}
{Bolton}, A.~S., {Burles}, S., {Schlegel}, D.~J., {Eisenstein}, D.~J., \&
  {Brinkmann}, J. 2004, \aj, 127, 1860

\bibitem[{{Bolton} {et~al.}(2008)}]{bol08}
{Bolton}, A.~S., {et~al.} 2008, \apj, 682, 964

\bibitem[{{Colless} {et~al.}(2001)}]{col_2df}
{Colless}, M., {et~al.} 2001, \mnras, 328, 1039

\bibitem[{{Cowie} {et~al.}(1996){Cowie}, {Songaila}, {Hu}, \& {Cohen}}]{cow96}
{Cowie}, L.~L., {Songaila}, A., {Hu}, E.~M., \& {Cohen}, J.~G. 1996, \aj, 112,
  839

\bibitem[{{Cramer} {et~al.}(2009){Cramer}, {Brown}, {Caldwell}, {Dupree},
  {Korzennik}, {Lykke}, \& {Szentgyorgyi}}]{cra09}
{Cramer}, C.~E., {Brown}, S., {Caldwell}, N., {Dupree}, A.~K., {Korzennik},
  S.~G., {Lykke}, K.~R., \& {Szentgyorgyi}, A. 2009, in Conference on Lasers
  and Electro-Optics/International Quantum Electronics Conference, {Optical
  Society of America Technical Digest}, JThE85

\bibitem[{{Cui} {et~al.}(2008){Cui}, {Ye}, \& {Bai}}]{cui08}
{Cui}, B., {Ye}, Z.~F., \& {Bai}, Z.~R. 2008, Acta Astronomica Sinica, 49, 327

\bibitem[{{Cuillandre} {et~al.}(1994){Cuillandre}, {Fort}, {Picat}, {Soucail},
  {Altieri}, {Beigbeder}, {Duplin}, {Pourthie}, \& {Ratier}}]{cuil94}
{Cuillandre}, J.~C., {et~al.} 1994, \aap, 281, 603

\bibitem[{{Cushing} {et~al.}(2004){Cushing}, {Vacca}, \& {Rayner}}]{cush04}
{Cushing}, M.~C., {Vacca}, W.~D., \& {Rayner}, J.~T. 2004, \pasp, 116, 362

\bibitem[{{Davis} {et~al.}(2003)}]{dav03}
{Davis}, M., {et~al.} 2003, in Society of Photo-Optical Instrumentation
  Engineers (SPIE) Conference Series, ed. {P.~Guhathakurta}, Vol. 4834,
  161--172

\bibitem[{{Dixon} {et~al.}(2007)}]{dix07}
{Dixon}, W.~V., {et~al.} 2007, \pasp, 119, 527

\bibitem[{{Glazebrook} \& {Bland-Hawthorn}(2001)}]{glaz01}
{Glazebrook}, K., \& {Bland-Hawthorn}, J. 2001, \pasp, 113, 197

\bibitem[{{Hall} {et~al.}(1994){Hall}, {Fulton}, {Huenemoerder}, {Welty}, \&
  {Neff}}]{hall94}
{Hall}, J.~C., {Fulton}, E.~E., {Huenemoerder}, D.~P., {Welty}, A.~D., \&
  {Neff}, J.~E. 1994, \pasp, 106, 315

\bibitem[{{Hewett} {et~al.}(1985){Hewett}, {Irwin}, {Bunclark}, {Bridgeland},
  {Kibblewhite}, {He}, \& {Smith}}]{hewett_85}
{Hewett}, P.~C., {Irwin}, M.~J., {Bunclark}, P., {Bridgeland}, M.~T.,
  {Kibblewhite}, E.~J., {He}, X.~T., \& {Smith}, M.~G. 1985, \mnras, 213, 971

\bibitem[{{Horne}(1986)}]{horne_86}
{Horne}, K. 1986, \pasp, 98, 609

\bibitem[{{Kelson}(2003)}]{kel03}
{Kelson}, D.~D. 2003, \pasp, 115, 688

\bibitem[{{Kinney} {et~al.}(1991){Kinney}, {Bohlin}, \& {Neill}}]{kin91}
{Kinney}, A.~L., {Bohlin}, R.~C., \& {Neill}, J.~D. 1991, \pasp, 103, 694

\bibitem[{{Le F{\`e}vre} {et~al.}(2005)}]{lef05}
{Le F{\`e}vre}, O., {et~al.} 2005, \aap, 439, 845

\bibitem[{{Marsh}(1989)}]{mar89}
{Marsh}, T.~R. 1989, \pasp, 101, 1032

\bibitem[{{Piskunov} \& {Valenti}(2002)}]{pisk02}
{Piskunov}, N.~E., \& {Valenti}, J.~A. 2002, \aap, 385, 1095

\bibitem[{{Robertson}(1986)}]{rob86}
{Robertson}, J.~G. 1986, \pasp, 98, 1220

\bibitem[{{Schlegel} {et~al.}(2009{\natexlab{a}}){Schlegel}, {White}, \&
  {Eisenstein}}]{sch_boss}
{Schlegel}, D., {White}, M., \& {Eisenstein}, D. 2009{\natexlab{a}}, in
  Astro2010: The Astronomy and Astrophysics Decadal Survey, 314

\bibitem[{{Schlegel} {et~al.}(2009{\natexlab{b}})}]{sch_bigboss}
{Schlegel}, D.~J., {et~al.} 2009{\natexlab{b}}, ArXiv e-prints
  (arXiv:0904:0468)

\bibitem[{{Shectman} {et~al.}(1996){Shectman}, {Landy}, {Oemler}, {Tucker},
  {Lin}, {Kirshner}, \& {Schechter}}]{schech_lcrs}
{Shectman}, S.~A., {Landy}, S.~D., {Oemler}, A., {Tucker}, D.~L., {Lin}, H.,
  {Kirshner}, R.~P., \& {Schechter}, P.~L. 1996, \apj, 470, 172

\bibitem[{{Steidel} {et~al.}(2003){Steidel}, {Adelberger}, {Shapley},
  {Pettini}, {Dickinson}, \& {Giavalisco}}]{steidel03}
{Steidel}, C.~C., {Adelberger}, K.~L., {Shapley}, A.~E., {Pettini}, M.,
  {Dickinson}, M., \& {Giavalisco}, M. 2003, \apj, 592, 728

\bibitem[{{Stubbs} \& {Tonry}(2006)}]{st06}
{Stubbs}, C.~W., \& {Tonry}, J.~L. 2006, \apj, 646, 1436

\bibitem[{{Wild} \& {Hewett}(2005)}]{wild05}
{Wild}, V., \& {Hewett}, P.~C. 2005, \mnras, 358, 1083

\bibitem[{{Willis} {et~al.}(2005){Willis}, {Hewett}, \& {Warren}}]{wil05}
{Willis}, J.~P., {Hewett}, P.~C., \& {Warren}, S.~J. 2005, \mnras, 363, 1369

\bibitem[{{Willis} {et~al.}(2006){Willis}, {Hewett}, {Warren}, {Dye}, \&
  {Maddox}}]{wil06}
{Willis}, J.~P., {Hewett}, P.~C., {Warren}, S.~J., {Dye}, S., \& {Maddox}, N.
  2006, \mnras, 369, 1521

\bibitem[{{York} {et~al.}(2000)}]{york_sdss}
{York}, D.~G., {et~al.} 2000, \aj, 120, 1579

\bibitem[{{Zanichelli} {et~al.}(2005)}]{zan05}
{Zanichelli}, A., {et~al.} 2005, \pasp, 117, 1271

\end{thebibliography}

\end{document}